\documentstyle[12pt]{article}
\begin{document}
\baselineskip 10mm


\vskip 4mm

\centerline{\large \bf Metastability of the Three-Dimensional Carbon Cluster
Prismane C$_8$}

\vskip 2mm

\centerline{V. F. Elesin, A. I. Podlivaev, L. A. Openov}

\vskip 2mm

\centerline{\it Moscow State Engineering Physics Institute
(Technical University),}
\centerline{\it Kashirskoe sh. 31, Moscow 115409, Russia}

\vskip 4mm

\begin{quotation}

Stability of recently predicted cage-like carbon
cluster C$_8$ (prismane) against its transformations to structures lying at
lower energies is studied theoretically. The smallest energy barrier
inhibiting prismane transformation is shown to be 0.44 eV, in qualitative
agreement with the value of activation energy for prismane decay calculated
by molecular dynamics simulations. A rather high value of the barrier
suggests the possibility of experimental observation of this cluster at
ambient pressure.

\end{quotation}

\vskip 6mm

{\bf 1. Introduction}

\vskip 2mm

Since fullerenes C$_{60}$ have been discovered \cite{Kroto} and synthesized
in macroscopic quantities \cite{Kratschmer}, there is a growing interest in
carbon clusters and nanostructures \cite{Hebard,Iijima,Piscoti}. This
interest has both fundamental and applied aspects. On the one hand, small
carbon clusters composed of $N \sim 10 - 100$ atoms differ from macroscopic
samples (and even from nanoparticles composed of $N \sim 10^4 - 10^5$ atoms)
in that there are no "inner" atoms in such clusters, i.e. all atoms
constituting the cluster lie on its "surface". A principal impossibility to
subdivide interatomic bonding into bulk and surface parts results in
essentially new physics and makes small clusters extremely interesting
objects for basic research \cite{Eaton,Weltner}. On the other hand, some
unusual properties of small carbon clusters can be used for practical
purposes. For example, metastable cubane C$_8$H$_8$ is considered as a
high-energy-density material that can store a considerable amount of energy
\cite{Eaton}; there is a hope that doping of C$_{36}$ fullerene can result in
its phase transition to a superconducting state with an anomalously high
critical temperature $T_c$ enhanced over the alkali-intercalated C$_{60}$
compounds \cite{Cote}, etc.

In order the small clusters could form bulk structures by means of
intercluster bonding, they should be {\it three-dimensional}.
Three-dimensional carbon clusters (cages) are believed to be stable for
$N\geq 20$ only \cite{Tomanek,Jones,Xu1}, while for clusters with $N<20$ the
stable structures are either one-dimensional (chains) or two-dimensional
(rings) \cite{Weltner,Tomanek,Xu1}. However, for $N<20$, carbon clusters can
exist in three-dimensional {\it metastable} states whose binding energies are
higher than the binding energy of the stable cluster with the same $N$ (for
example, the occurrence of a metastable C$_{14}$ cage has been reported in
Ref. \cite{Jones}, based on the results of density functional calculations).

Several cage-like metastable isomers C$_8$ have been proposed in
Refs. \cite{Kobayashi,Jones2,Openov} using the Harris-functional
approximation combined with a generalized-gradient approximation
\cite{Kobayashi}, the local spin density and gradient-corrected
approximations to the exchange-correlation energy \cite{Jones2} as well as
the trasferable tight-binding potential \cite{Openov}. As far as we know,
eight-atom isomers have not been observed experimentally yet, they are the
smallest three-dimensional carbon clusters found so far theoretically. The
key issue for a possibility of experimental observation of such clusters is
their relative stability. Note, however, that the stability of isomers C$_8$
have not been quantified in refs. \cite{Kobayashi,Jones2}, e.g., the
transformations to structures lying at lower energies
have not been studied. Meanwhile, the results of {\it ab initio}
calculations are known to depend strongly on the way that electron
correlations are treated \cite{Weltner,Jones2}, and hence a conclusion
about the cluster metastability may appear to be just an artifact of
approximations used.

Recently we have predicted \cite{Openov} the existence of an eight-atom
metastable cage-like cluster C$_8$ that has the shape of a six-atom
triangular prism with two excess atoms above and below its bases. We gave
this cluster the name "prismane". It is shown in Fig.1.
The binding energy of prismane is 0.45 eV/atom higher than
the binding energy of the stable one-dimensional eight-atom cluster shown in
Fig.2. Nevertheless, molecular dynamics simulations gave evidence for a
relatively high stability of prismane, the activation energy for prismane
decay was estimated to be about 1 eV \cite{Openov}, suggesting that
prismane lifetime is rather long and that this cluster may be observed
experimentally.

However, although the finite-temperature molecular dynamics simulations
provide the most direct way for evaluation of the cluster lifetime, such
calculations are extremely time-consuming since the process of cluster decay
is probabilistic in nature, and hence a huge statistics should be accumulated
in order to draw a definite conclusion about the values of the cluster
lifetime and activation energy. Alternatively, useful complementary
information concerning characteristics of the metastable state can be
provided by calculations of the heights of energy barriers inhibiting
spontaneous transformation of the cluster to the lower-energy atomic
configuration. Such calculations have been carried out, e.g., to study
the atmospheric-pressure stability of energetic phases of carbon
\cite{Mailhiot1} and polymeric nitrogen \cite{Mailhiot2}. It was the purpose
of this work to find the minimum energy barrier separating the metastable
prismane structure C$_8$ shown in Fig.1 from the stable chain structure
in order to see if such a cluster can be observed experimentally at
non-exotic conditions.

\vskip 6mm

{\bf 2. Computational details}

\vskip 2mm

To calculate the binding energy of a cluster having an arbitrary atomic
configuration, we have used a transferable tight-binding potential
\cite{Xu1,Xu2} that had been proven to reproduce accurately the
energy-versus-volume diagram of carbon polytypes and to give a good
description of both small clusters and bulk structures of carbon
\cite{Xu1,Xu2}. On the one hand, this potential describes the structure and
energetics of small carbon clusters quite well, the difference in bond
lengths and binding energies between our results and available
{\it ab initio} calculations \cite{Weltner} usually did not exceed 10$\%$. On
the other hand, the technique used greatly simplifies a reliable
evaluation of characteristics of metastable states as compared with
{\it ab initio} approaches \cite{Lee,Dunn}.

The search for the minimum energy barrier separating two locally stable
atomic configurations is equivalent to the search for a saddle point in the
total-energy surface as a function of the atomic coordinates. Such a saddle
point corresponds to the unstable configuration of atoms in the cluster. The
algorithm of the search for a saddle point is based on a continuous
transformation of the cluster along the directions of atomic motion in the
softest oscillation mode of the cluster, i.e. the mode with the minimum
oscillation frequency. This algorithm is analogous to the "soft-mode walking"
procedure \cite{Simons} in which the information regarding both gradients and
Hessian (energy curvatures) is used to systematically proceed up the bottom
of a valley (local minimum) to a transition (saddle) state. In uphill walks
from the minimum-energy geometry of the cluster to the transition state, the
energy is maximized in the soft-mode direction and minimized in all
orthogonal directions, thereby forcing the walk to be in a stream bed
\cite{Simons}. Apart from the efficient search for a saddle point, such an
algorithm gives the energy spectrum of mechanical oscillations of the
cluster. The knowledge of this spectrum allows for a reliable discrimination
between the metastable and long-living unstable states of the cluster
(sometimes this cannot be done by means of molecular dynamics simulations).

All calculations were carried out at pressure $P=0$, i.e. in fact at
ambient (atmospheric) pressure \cite{Mailhiot2}.

\vskip 6mm

{\bf 3. Results and discussion}

\vskip 2mm

We define the binding energy $E_b$ of a C$_8$ cluster as
\begin{equation}
E_b=E(8)-8E(1),
\label{Eb}
\end{equation}
i.e. as the difference between the total energy of the cluster, $E(8)$, and
the energy $8E(1)$ of eight isolated carbon atoms. The binding energy of the
metastable prismane (P) is $E_b^P=-40.80$ eV. The binding energy of the
stable eight-atom chain (C) is $E_b^C=-44.37$ eV.

Starting with the prismane structure, we first computed the energy spectrum
of its mechanical oscillations. We have found that all oscillation
frequencies are real, giving evidence that the prismane structure corresponds
to a local minimum in the total-energy surface as a function of the atomic
coordinates, i.e. the prismane is indeed a metastable cluster.

We now consider the energetics associated with the prismane transformation to
the stable chain configuration in the soft-mode direction. Fig.3 shows the
difference $\Delta E_b=E_b-E_b^C$ between the binding energy of the
transformed C$_8$ cluster, $E_b$, and $E_b^C$ versus the generalized
transformation coordinate $R$ defined as a mean square deviation of atomic
coordinates $x_i$, $y_i$, $z_i$ of the transformed cluster from their
corresponding values $x_{i0}$, $y_{i0}$, $z_{i0}$ in the prismane structure
($i=1-8$ is the number of atom in the cluster):
\begin{equation}
R^2=\sum_{i=1}^{8}[(x_i-x_{i0})^2+(y_i-y_{i0})^2+(z_i-z_{i0})^2].
\label{R}
\end{equation}
One can see that the function $\Delta E_b(R)$ has a local minimum at $R=0$
corresponding to the metastable prismane structure. The value of
$\Delta E_b(R)$ increases with $R$ and passes through the maximum S1 at
$R\approx 0.5$ {\AA}. This maximum corresponds to the unstable configuration
"scorpion" shown in Fig.4. The binding energy of the "scorpion" is
$E_b^{S1}=-40.36$ eV, so that the energy barrier inhibiting spontaneous
prismane transformation to the lower-energy atomic configuration appears to
be $U=E_b^{S1}-E_b^P=0.44$ eV.

We have found (see Fig.3) that further increase in $R$ leads to the prismane
transformation at $R\approx 6.5$ {\AA} not into the stable eight-atom chain
(as one could expect) but into another metastable configuration "frying pan"
(FP) shown in Fig.5. This configuration corresponds to the local minimum in
the total-energy surface. The binding energy of the "frying pan"
$E_b^{FP}=-42.85$ eV is 2.05 eV below the binding energy $E_b^P$ of the
prismane but 1.52 eV higher than that of the stable chain, $E_b^C$.

Next we have studied the transformation path of the "frying pan" in its
soft-mode direction. Fig.6 shows the value of $\Delta E_b=E_b-E_b^C$ as a
function of the transformation coordinate $R$ defined by Eq.(\ref{R}), where
$x_{i0}$, $y_{i0}$, $z_{i0}$ are now the atomic coordinates of the "frying
pan". One can see that $\Delta E_b(R)$ first increases with $R$, passes
through the maximum S2 at $R\approx 2.7$ {\AA} and then decreases down to
zero at $R\approx 6$ {\AA}, so that the stable chain structure (see Fig.2) is
finally reached. The maximum of $\Delta E_b(R)$ corresponds to the unstable
configuration shown in Fig.7. The binding energy of this configuration is
$E_b^{S2}=-42.08$ eV, so that the energy barrier inhibiting the "frying pan"
transformation to the linear chain is equal to $U_1=E_b^{S2}-E_b^{FP}=0.77$
eV. Hence, the prismane transformation into the stable eight-atom chain is a
two-step process that takes place via an "intermediary" metastable
configuration "frying pan".

Note that there also exist the paths of prismane transformation other than
that shown in Figs. 3 and 6. The latter, however, is characterized by the
smallest energy barrier $U$ inhibiting transformation of the prismane
structure to the lower-energy configurations. It is the path with the
smallest barrier that is key to the issue of metastability.

We stress that the data of finite-temperature molecular-dynamics simulation
\cite{Openov} are consistent with the results presented above. For all
temperatures studied, 800 K $< T <$ 1800 K, the prismane first transforms to
the "frying pan" configuration, and next to the linear chain. The activation
energy $E_a=0.82$ eV for prismane decay estimated in \cite{Openov} on the
basis of the temperature dependence of prismane lifetime is about twice the
height of the smallest energy barrier $U=0.44$ eV calculated in this work.
Note, however, that the cluster lifetime depends, in general, on the full
topology of the total-energy surface, beyond just the minimum energy barrier.

\vskip 6mm

{\bf 4. Conclusions}

\vskip 2mm

We have numerically examined the metastability of recently predicted
cage-like carbon cluster prismane C$_8$. It
was demonstrated that the minimum energy barrier separating the prismane
structure from the lower-energy atomic configuration amounts to 0.44 eV.
Combined with the data on finite-temperature molecular-dynamics simulation of
prismane decay, our result points to rather long lifetime of the prismane in
its metastable state, suggesting the possibility of experimental synthesis of
this cluster.

\vskip 6mm

{\bf Acknowledgments}

\vskip 2mm

The work was supported by the Contract DSWA01-98-C-0001 and by the Russian
Federal Program "Integration", projects No A0133 and A0155.


\newpage
\centerline{\bf Figure captions}
\vskip 2mm

Fig.1. Metastable prismane C$_8$. Binding energy $E_b^P=-40.80$ eV. Bond
lengths: $d_{AB}=2.31$ {\AA}, $d_{AC}=1.28$ {\AA}, $d_{AD}=1.47$ {\AA}. Bond
angles: $\angle BAC = 90^o$, $\angle ADB = 104^o$, $\angle ABE = 60^o$.

Fig.2. Stable chain C$_8$. Binding energy $E_b^C=-44.37$ eV.

Fig.3. Calculated difference $\Delta E_b=E_b-E_b^C$ between the binding
energy $E_b$ of the C$_8$ cluster transformed from the prismane structure (P)
in the soft-mode direction and the binding energy $E_b^C$ of the stable
eight-atom chain as a function of the transformation coordinate $R$. The
prismane structure corresponds to $R=0$. The saddle point configuration
"scorpion" (S1) is realized at $R\approx 0.5$ {\AA}. The lower-energy
metastable configuration "frying pan" (FP) is reached at $R\approx 6.5$
{\AA}. The energy barrier for the prismane $\rightarrow$ "frying pan"
transformation is $U=E_b^{S1}-E_b^P=0.44$ eV.

Fig.4. Unstable configuration "scorpion" (S1 in Fig.3). Binding energy
$E_b^{S1}=-40.36$ eV.

Fig.5. Metastable configuration "frying pan" (FP in Figs. 3 and 6). Binding
energy $E_b^{FP}=-42.85$ eV.

Fig.6. Calculated difference $\Delta E_b=E_b-E_b^C$ between the binding
energy $E_b$ of the C$_8$ cluster transformed from the "frying pan" structure
(FP) in the soft-mode direction and the binding energy $E_b^C$ of the stable
eight-atom chain as a function of the transformation coordinate $R$. The
"frying pan" structure corresponds to $R=0$. The saddle point configuration
(S2) is realized at $R\approx 2.7$ {\AA}. The stable chain configuration (C)
is reached at $R\approx 6$ {\AA}. The energy barrier for the "frying pan"
$\rightarrow$ chain transformation is $U_1=E_b^{S2}-E_b^{FP}=0.77$ eV.

Fig.7. Unstable configuration S2 (see Fig.6). Binding energy
$E_b^{S2}=-42.08$ eV.

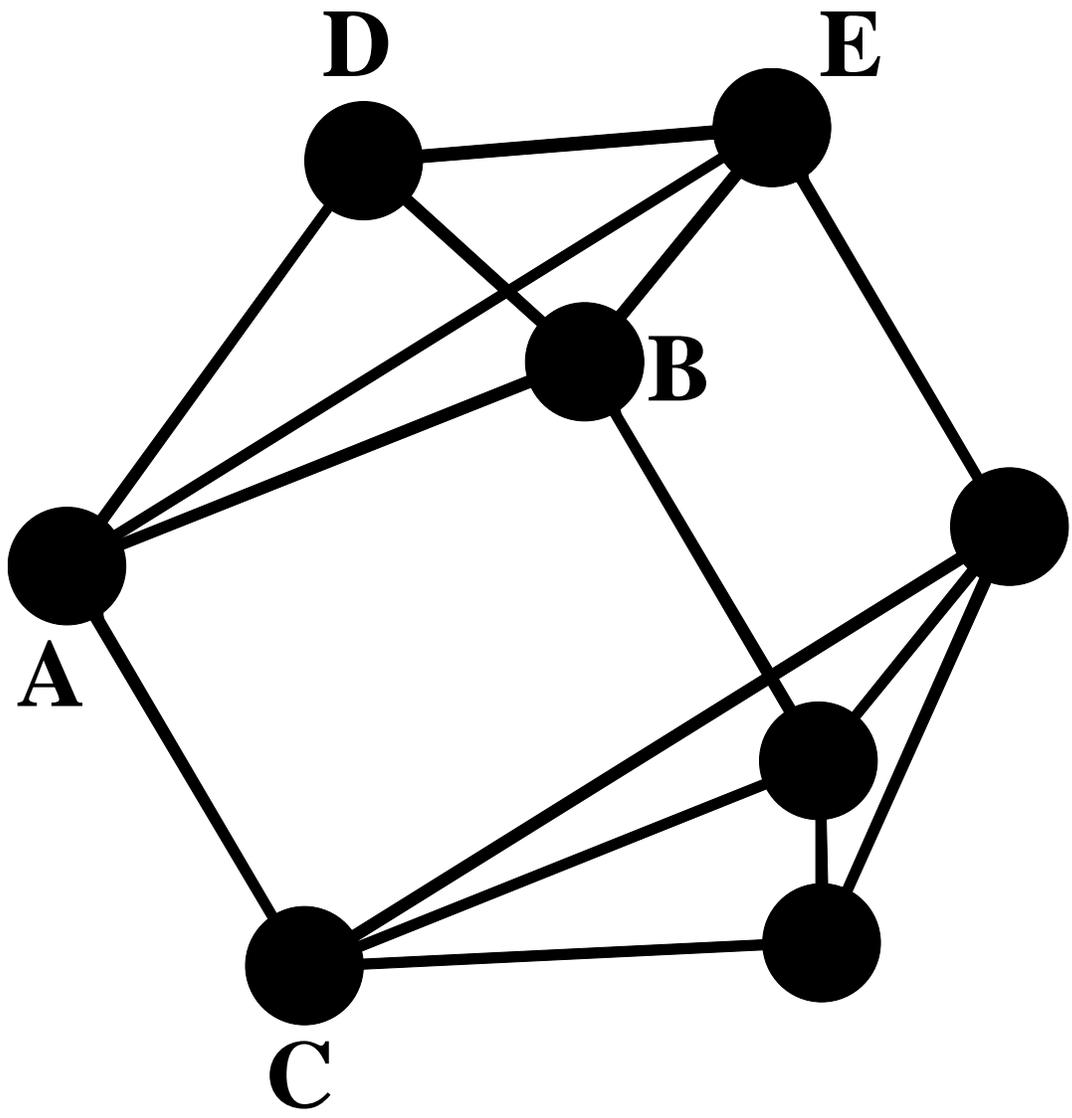
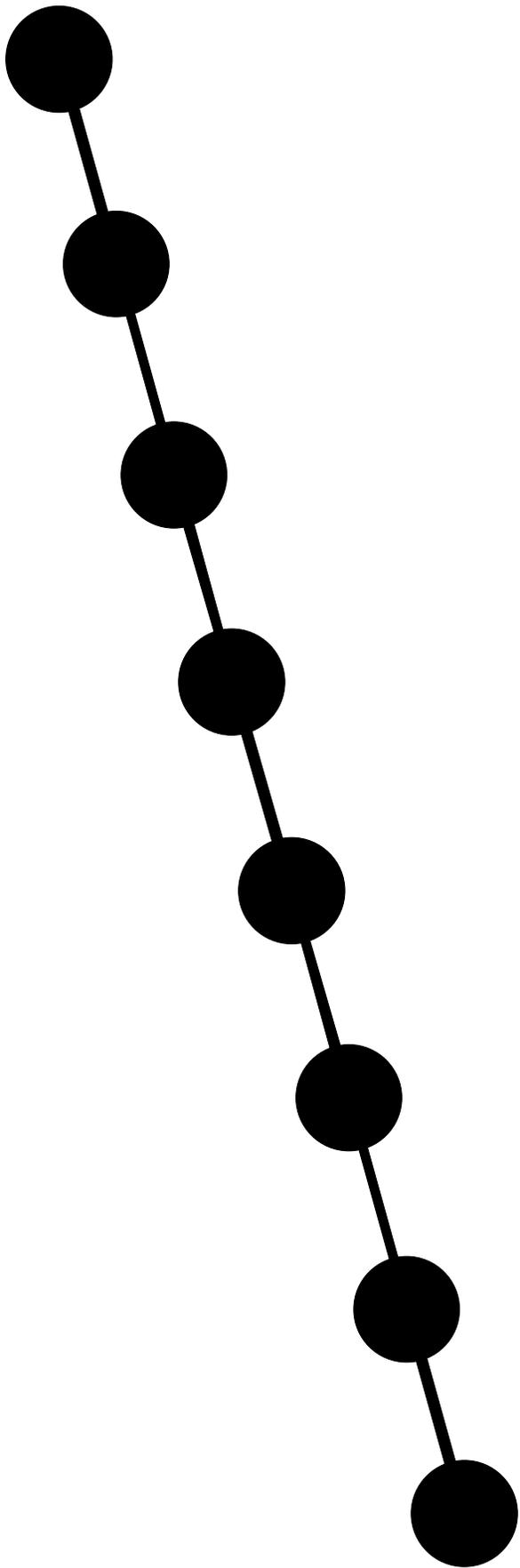
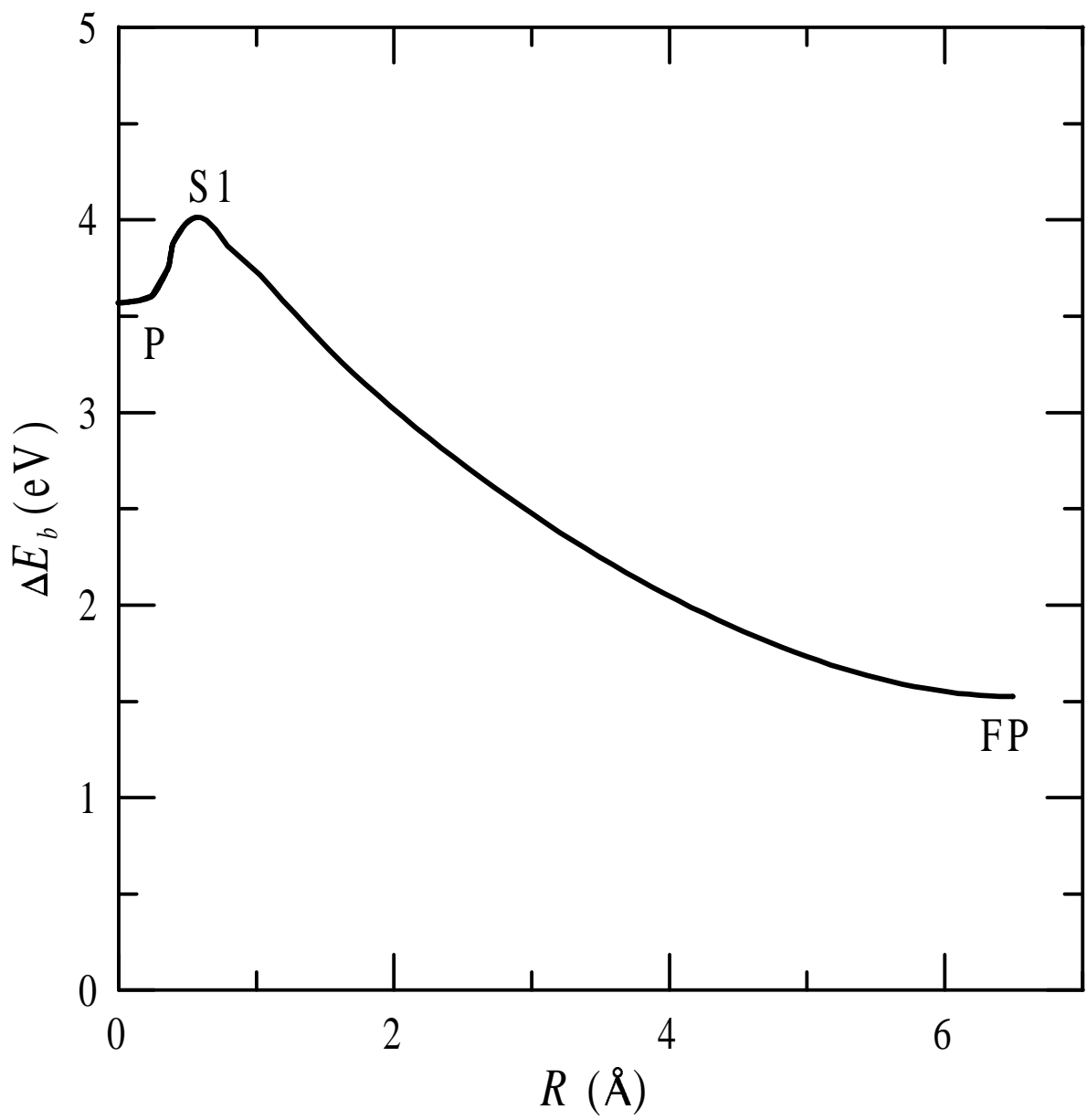
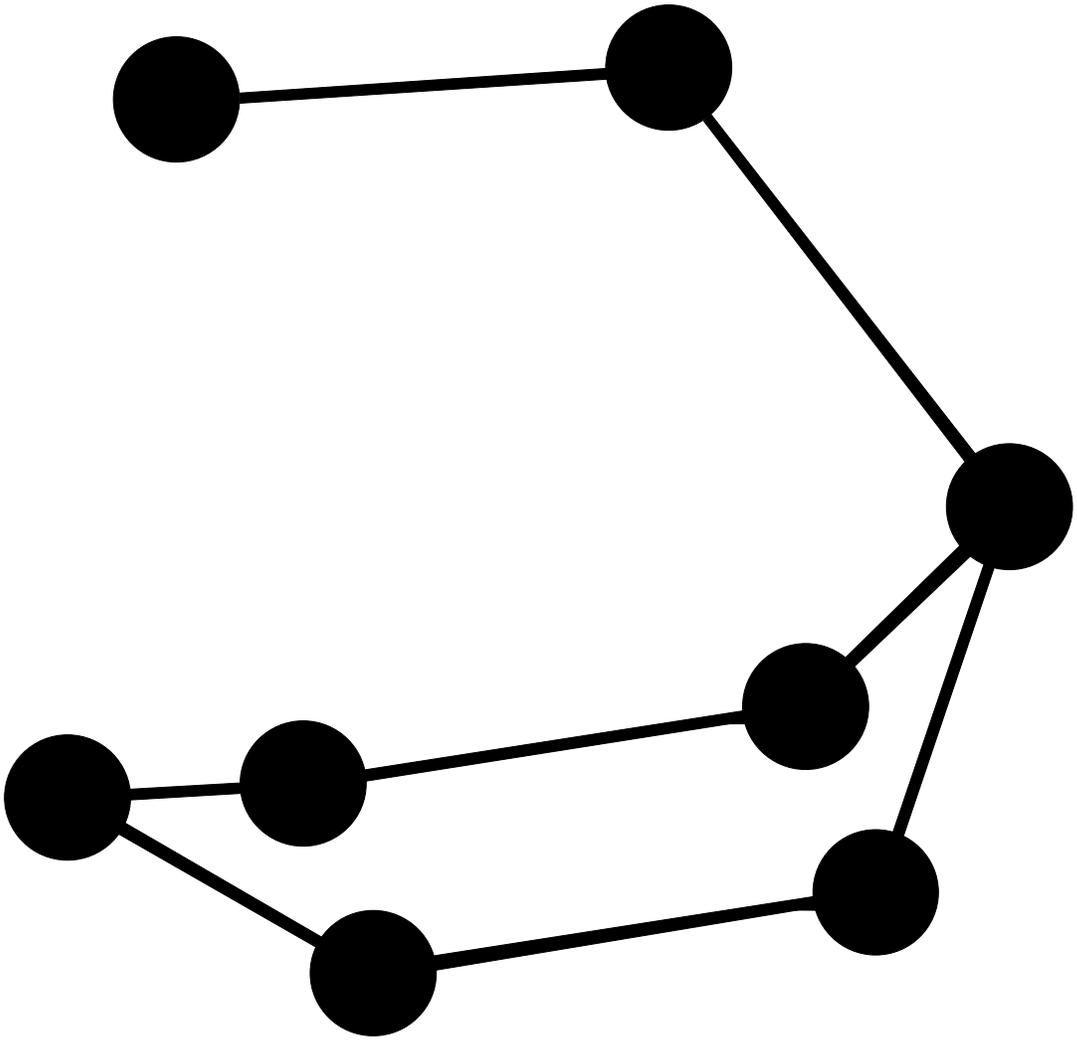
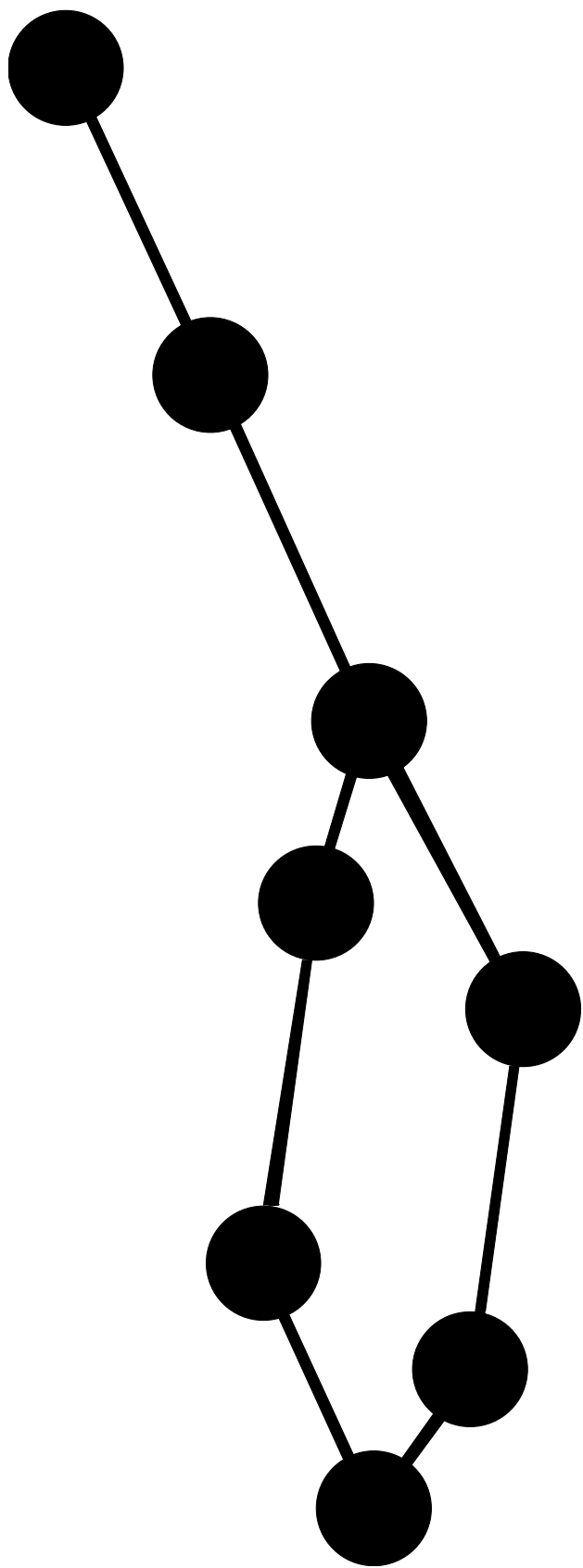
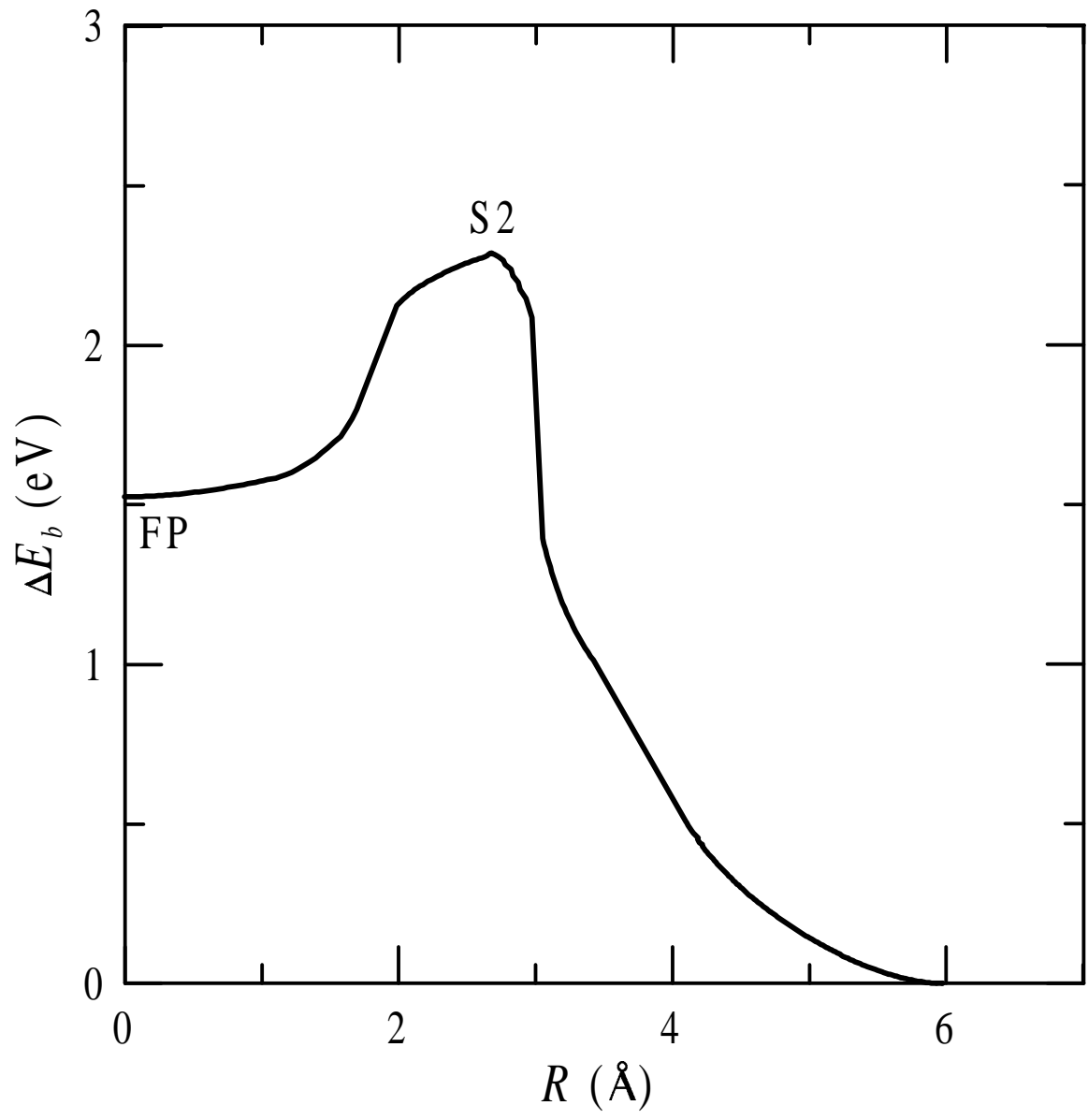
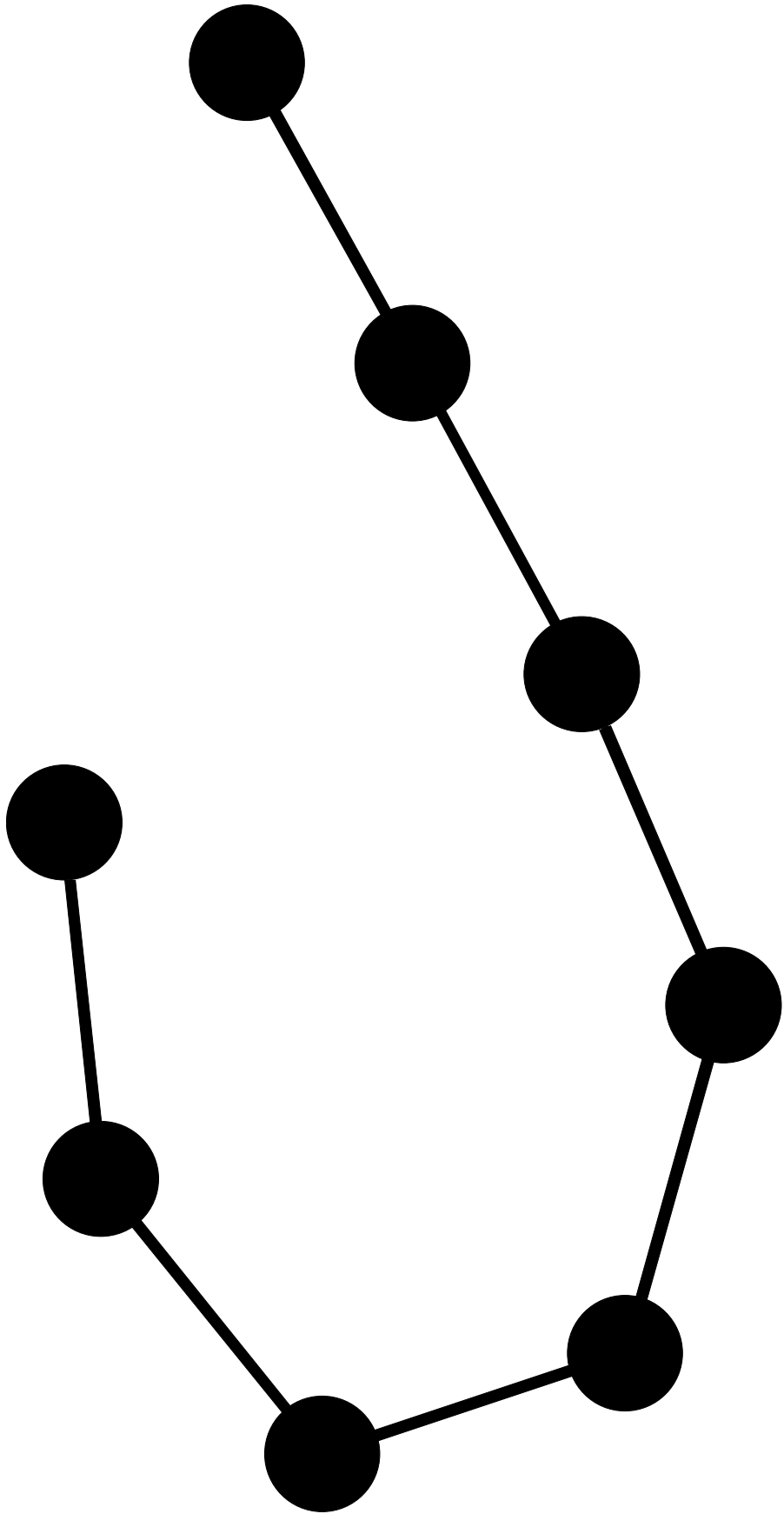
\end{document}